# Cells Solved the Gibbs Paradox by Learning to Contain Entropic Forces

Josh E. Baker[1*]

[1]University of Nevada, Reno School of Medicine: Reno, NV 89521 USA

*Corresponding author. Email: jebaker@unr.edu

**Abstract:** As Nature's version of machine learning, evolution has solved many extraordinarily complex problems, none perhaps more remarkable than learning to harness an increase in chemical entropy (disorder) to generate directed chemical forces (order). Using muscle as a model system, here I unpack the basic mechanism by which life creates order from disorder. In short, evolution tuned the physical properties of certain proteins to contain changes in chemical entropy. As it happens these are the "sensible" properties Gibbs postulated were needed to solve his paradox.

**Summary:** A binary mechanical system provides a unifying description of molecular mechanics and emergent thermodynamics and provides an explicit solution to the Gibbs paradox.

**Introduction**

In 1876 J.W. Gibbs identified a paradox in his chemical thermodynamic treatment of entropy that has confounded scientists from Boltzmann to Einstein and that remains an intriguing puzzle to this day(*1*). There is presently no one explicit solution to the paradox, and it has been suggested that "the multiplicity of solutions proposed… [implies] that there are different ways of conceiving the foundations of thermodynamics"(*1*). Biological systems that have evolved to contain entropic forces(*2, 3*) provide a model system for studying this paradox, which I use here to show that the multiplicity of proposals are not distinct concepts but rather elements of a single explicit solution.

**The Paradox Applied to a Two-State Chemical Reaction**

A version of the Gibbs paradox is illustrated in Fig. 1. Figure 1A is a kinetic scheme for a chemical reaction in which a molecule reversibly isomerizes between two chemical states, B and Y, differing only in color. In state B, the molecule is blue, and in state Y, the molecule is yellow. The molecule switches between these states with forward, $f_+$, and reverse, $f_-$, rates. If at time $t = 0$ a system contains 10 such molecules all in state B (Fig. 1B, left), then at a later time $t > \tau = \frac{1}{f_+ + f_-}$ the system will equilibrate with molecules distributed (equally if $f_+ = f_-$) between states B and Y (Fig. 1B, right). In a solution containing many molecules, this reaction appears as a blue solution that irreversibly turns green (Fig. 1C, left to right).

Figure 1C resembles experiments in which two drops of different colored dyes are placed into a glass of water and mix spontaneously and irreversibly through diffusion; only here the spontaneous change in color occurs through a two-state chemical reaction. Because in both cases, an irreversible mixing of colors is energetically driven by an increase in system entropy, here I refer to the equilibration of the chemical reaction in Fig. 1C as "mixing".

The spontaneous change in color in Fig. 1C is energetically driven by the entropic contribution, ΔS, to the free energy for the reaction in Fig. 1A, where ΔS is defined independent of the colors of the two states so long as the difference, $d$, between them (here a wavelength) is distinguishable. When the two states become indistinguishable ($d = 0$), the reaction no longer occurs because there is only one state (one color). At this point ΔS abruptly vanishes. The paradox is that a subtle change in the difference between states (from $d$ being barely detectable to $d = 0$) has unexplained, discontinuous energetic consequences.

Most proposed solutions to this paradox are based on arguments invoking a mutable ΔS(*1*). Maxwell argued that ΔS is defined by the mind that perceives molecular differences (e.g., Maxwell's demon). Gibbs argued that ΔS is only defined by sensible properties. Not surprisingly, Planck argued that ΔS requires finite differences, $d_{crit}$, between molecular states, claiming "Chemical differences between… two substances in general cannot be represented by a continuously variable quality; and that we instead have to do with discrete distinctions… This circumstance creates a principal opposition between chemical and physical properties since the latter must always be regarded as continuously variable"(*1*).

The common assumption, made here by Planck and elsewhere by others, that entropic changes must be continuously variable is the basis for the infamous arbitrary division by N! employed by Boltzmann in his analysis of the Gibbs paradox(*1*). However, neither chemical nor physical properties of a chemical reaction are continuous, and as shown here, by considering the discrete changes in system entropy associated with discrete chemical steps, Boltzmann's N! term cancels, and both Gibbs' sensible properties and Planck's $d_{crit}$ are explicitly defined.

Biological systems like muscle have evolved to contain entropic forces within cells by tuning proteins to optimize their sensible properties and $d_{crit}$; as such, they serve as model systems for formally developing these concepts. Here, I define the energy of mixing for the reaction in Fig. 1. Next, I describe a mechanism for un-mixing (a mechanistic difference, $d$, between states) inspired by the chemistry of muscle contraction. Finally, I calculate the $d$-dependent energy required for un-mixing, providing a unifying description of molecular mechanics and emergent thermodynamics.

**The Energy of Mixing.**

According to Boltzmann, the entropy, S, of a system is $k_B \ln \Omega$, where $\Omega$ is the number of microstates accessible to the system. Within a given state, [$N_B$, $N_Y$], of the system in Fig. 1B, the number of microstates is $\Omega = \frac{N!}{N_B! N_Y!}$, where $N_B$ and $N_Y$ are the number of molecules in states B and Y, and $N = N_B + N_Y$. With a single chemical step from blue to yellow, the number of microstates within this new state [$N_B–1, N_Y+1$] becomes $\Omega = \frac{N!}{(N_B-1)!(N_Y+1)!}$. The change in system entropy, $\Delta S$, with a chemical step from [$N_B, N_Y$] to [$\underline{N_B}–1, \underline{N_Y}+1$] is $k_B \ln \frac{(N_B-1)!(N_Y+1)!}{(N_B)!(N_Y)!}$ (note the N! terms cancel), and according to Boltzmann

$$\Delta S = k_B \ln \frac{N_Y+1}{N_B}.$$

According to Gibbs, the entropic contribution to the free energy that drives the mixing reaction (Fig. 1) is then

$$T\Delta S = k_B T \ln \frac{N_Y + 1}{N_B}. \quad \text{(Eq. 1)}$$

Because a color change has little physical impact, here I consider a version of the two-state scheme in Fig. 1 in which the difference between states is mechanical. Specifically, I consider a two-state chemical reaction where the difference between states is a measurable displacement, $d$ (Fig. 2A).

**A Binary Mechanical System**

Figure 2 describes a binary mechanical system that accounts for many mechanical, chemical and energetic aspects of muscle contraction (*3, 4*). In Fig. 2A, actin filament binding induces a conformational change in myosin (a structural lever arm rotation) that displaces the actin filament a distance $d$ (*5, 6*). For continuity with Fig. 1, a hypothetical fluorophore bound to myosin changes color from blue to yellow when myosin binds actin (Fig. 2A). Focusing on entropic forces, here I assume that the actin-myosin binding free energy, $\Delta G°$, is zero (i.e., $f_+ = f_–$).

Figure 2B is the same mixing reaction illustrated in Fig. 1B, only here myosin molecules that are attached to a fixed surface move an actin filament attached to a moveable surface a distance, $d$, with each discrete chemical step from B to Y. In other words, an increase in system entropy generates directed movement. This entropically-driven contraction of the system can be reversed by physically pulling on the system to expand it. The change in external force, $\Delta F_{ext}$, required to mechanically pull the system from green (a mixture of yellow and blue) back to blue, can be calculated from changes in both molecular, $\Delta F_1$, and entropic, $\Delta F_S$, forces.

In single molecule mechanics studies, we have shown that a single chemical step from B to Y displaces a spring of stiffness $\kappa_{sys}$, generating force, $\kappa_{sys}d$,(6–8) where $d$ can be experimentally measured and controlled by genetically engineering different myosin lever arm lengths(9). We have also shown(6, 10) that with a chemical reversal of this step force decreases by

$$\Delta F_1 = -\kappa_{sys}d. \quad (Eq.\ 2)$$

A single system spring of stiffness $\kappa_{sys}$ provides a useful construct for uniting molecular force generation and system forces. As illustrated in Fig. 2C, one end of a system spring is extended or shortened by reversible chemical steps, $d$ (bottom), while the other end (top) of the spring equilibrates with a macroscopic (e.g., entropic) force(3).

When the system in Fig. 2B (right) is pulled to generate force $\Delta F_{ext} = -\Delta F_1$ (Fig. 2C, left), the system responds with a single molecule step from state Y to B (Fig. 2C, left to right) that reverses $\Delta F_{ext}$ (Eq. 2). This decrease in system force, $\Delta F_1$, with a single molecule step energetically drives the unmixing step. However, the system does not equilibrate with a single molecule step; it equilibrates with the chemical relaxation of the system. Upon equilibration the increase in entropy associated with a transition from states [5,5] to [4,6] is balanced against an increase in entropic force, $F$, that is defined by the equilibrium free energy equation for the reaction in Fig. 2A:

$$\Delta G° + T\Delta S + Fd = 0. \quad (Eq.\ 3)$$

Here $Fd$ is the work performed by a single step $d$ against the system force, $F$, and $T\Delta S$ is defined by Eq. 1. Assuming $\Delta G° = 0$, the equilibrium entropic force is

$$F = \frac{k_BT}{d}\ln\frac{N_B}{N_Y+1}. \quad (Eq.\ 4)$$

Consistent with Eq. 4, we have shown experimentally(11) that when a force, $F$, is applied to an equilibrium muscle system in which the actin-myosin binding affinity is chemically diminished, the observed distribution of states changes with $F$ as $\frac{N_Y}{N_B} = e^{-\frac{Fd}{k_BT}}$, demonstrating that, consistent with Eq. 4, an equilibrium mixture of force generating myosin molecules can be unmixed by increasing $F$. According to Eq. 4, the change in entropic force with a change in system state from $[N_B,N_Y]$ to $[N_B–1,N_Y+1]$ is

$$\Delta F_S = \frac{T\Delta\Delta S}{d} = \frac{k_B T}{d} \ln \frac{(N_B + 1)(N_Y+1)}{(N_B)(N_Y)} \quad \text{(Eq. 5)}$$

which ranges from zero when fully mixed to $\frac{k_B T}{d}\ln 2$ when fully unmixed.

According to continuous, near-equilibrium definitions of entropic changes, small external increments in the system force, $\Delta F_{ext} = \Delta F_S$ (Eq. 5), reverse the mixing reaction along a smooth isotherm (Eq. 4). However, in a discrete physical chemical analysis, a transient change in mechanical force, $-\Delta F_1$ (Eq. 2), physically drives the un-mixing step. Combined, the change in external force required to drive the un-mixing reaction forward, $-\Delta F_1$, against the increased entropic force, $\Delta F_S$, is $\Delta F_{ext} = \Delta F_S - \Delta F_1$, or

$$\Delta F_{ext} = \frac{T\Delta\Delta S}{d} + \kappa_{sys} d. \quad \text{(Eq. 6)}$$

Figure 3A illustrates this tripartite sequence of mechanochemical events for a system containing $N = 11$ molecules. When the system force is increased, $\Delta F_{ext}$, by externally pulling on the system (Fig. 3A, up arrow), the system responds with a chemical step from state [8,3] to [9,2], which occurs with both a decrease in molecular mechanical force, $-\kappa_{sys} d$ (Fig. 3A, blue arrow), and an increase in entropic force, $\frac{T\Delta\Delta S}{d}$ (Fig. 3A, red arrow), resulting in a new equilibrium force along the isotherm (Eq. 4, red line). The chemical reversal of the above process (Fig. 3A, gray arrows and text) defines a finite minimum work loop around a single chemical step.

The total driving force for un-mixing is $\kappa_{sys} d - T\Delta\Delta S/d$, which means that when $\kappa_{sys} d = T\Delta\Delta S/d$, un-mixing is physically not possible. This defines a finite minimum difference between states of

$$d = d_{crit} \equiv \sqrt{\frac{T\Delta\Delta S}{\kappa_{sys}}}. \quad \text{(Eq. 7)}$$

Equation 7 is more than simply an equilibrium condition. It describes the point at which a chemical equilibrium is unaffected by work performed on the system, $\Delta F_{ext}$. Beyond this point, when $\Delta F_{ext}$ exceeds that defined by Eq. 6, $\Delta F_{ext}$ is simply a passive force both uncoupled from chemistry (it has no effect on Eq. 7) and incapable of further unmixing the system. While pulling on the system harder to generate forces beyond $\Delta F_{ext}$ (Eq. 6) might forcibly detach molecules or even tear the system apart (chemically irreversible processes), the reversible un-mixing reaction is not mechanically driven by $\Delta F_{ext}$; it is mechanically driven by $-\kappa_{sys} d$, which is defined by finite molecular parameters. In other words, the finite molecular difference, $d_{crit}$, postulated by

Planck, is related to the sensible property, $-\kappa_{sys}d$, postulated by Gibbs through a discrete change in system entropy (Eq. 7).

Because $T\Delta\Delta S/d$ increases from 0 to $\frac{k_B T}{d}\ln 2$ with unmixing, Eq. 7 indicates that a reaction can be unmixed to some extent even with a relatively small $d$. This is illustrated in Fig. 3B where increments of $\Delta F_{ext}$ unmix the reaction along the isotherm (Eq. 4) until $d = d_{crit}$ (Fig. 3B, asterisk) beyond which point the reaction cannot physically be further unmixed.

When $d < d_{crit}$ entropic force dominates and mixing occurs spontaneously and unstoppably against a relatively small mechanical force, $\kappa_{sys}d$. At the other extreme, when $d \gg d_{crit}$, there is no chemical contribution to mixing or unmixing (Eq. 6), and at this molecular mechanical limit the reaction is driven forward and backward by external mechanical steps alone, $\Delta F_{ext} = \Delta F_1$.

**Conclusion**

The above analysis provides a solution to the Gibbs paradox as it pertains to a binary mechanical system (Fig. 2B). The analysis implies that only at the discrete finite limit of chemical steps can we define changes in both molecular and entropic forces (Fig. 3A) that together unify molecular mechanics (top descending limb) and emergent thermodynamics (bottom ascending limb). Only at this discrete limit can we define the molecular mechanical force, $-\kappa_{sys}d$, (Fig. 3A, negative slope) that drives a chemical step against the entropic force of mixing, $T\Delta\Delta S/d$, (Fig. 3A, positive slope). And only at this discrete limit do we recognize that un-mixing is physically not possible when $-\kappa_{sys}d$ (the driving mechanical force) is less than $T\Delta\Delta S/d$ (the resistive entropic force).

Equivalently, un-mixing is physically not possible when the mechanical energy, $-\kappa_{sys}d^2$, is less than the entropic energy, $T\Delta\Delta S$; as such $-\kappa_{sys}d^2$ can be viewed as a finite physical (sensible) container of $T\Delta\Delta S$. When the container is large, it can hold large amounts of $T\Delta\Delta S$. When the container is small, only small amounts of $T\Delta\Delta S$ can be held in a system with the excess irretrievably spilling out into the universe. In Fig. 3B, the maximum extent of unmixing changes with the size of the container (Fig. 3B, maroon bar). Here, the approach to indistinguishable states (as $d$ becomes small) is continuous. The container (the capacity to measure, use or reverse $T\Delta\Delta S$) becomes infinitesimally small ($\kappa_{sys}d^2$ gets small) as the two states become infinitesimally similar, and when $d$ becomes zero, there is at once both no container and nothing to contain.

Through all processes and at all scales across the universe entropy increases, and this increasing disorder can be locally ordered (measured, used, or reversed) only when placed in a proper container. The primordial soup consisted of chemical reactions dominated by thermal energy and increasing entropy, and despite the exacting physical relationships required (Eqs. 6 and 7), biological systems have evolved highly effective mechanisms for containing within cells the $T\Delta\Delta S$ for certain reactions. Thus, it is no surprise that the chemical reaction that drives muscle contraction informs us of these relationships.

Large containers ($d \gg d_{crit}$) that dominate entropy flip the agency of a reaction ($\Delta F_{ext} = \Delta F_1$, with no chemical forces). Because the primordial soup contained a paucity of directed external forces, $\Delta F_{ext}$, available to order cells but an abundance of increasing entropy, $T\Delta\Delta S/d$, available to be ordered by them, catabolic reactions evolved as unidirectional chemical forces (e.g., $\frac{T\Delta\Delta S}{d}$) that drive unidirectional changes in surrounding forces, $\Delta F_{ext}$ ($d \approx d_{crit}$) not the other way around ($d \gg d_{crit}$). This emergent perspective is the antithesis of the molecular (corpuscular) mechanic myth (*12*, *13*) that gears and springs from the primordial soup were pieced together using rational mechanics ($\Delta F_{ext} = \Delta F_1$). Paraphrasing Gibbs, we will never find in molecular biology an *a priori* foundation for the principles of biological function. The above thermodynamic relationships (Fig. 3A) transform our understanding of how muscle works(*3*, *4*) and have broad implications for both natural and synthetic biology.

**References and Notes**

1. O. Darrigol, The Gibbs paradox: Early history and solutions. *Entropy*. **20** (2018), doi:10.3390/e20060443.
2. M. S. Kellermayer, S. B. Smith, H. L. Granzier, C. Bustamante, Folding-Unfolding Transitions in Single Titin Molecules Characterized with Laser Tweezers. **276**, 1112–1116 (1997).
3. J. E. Baker, Thermodynamics and Kinetics of a Binary Mechanical System: Mechanisms of Muscle Contraction. *Langmuir*. **38**, 15905–15916 (2022).
4. J. E. Baker, D. D. Thomas, A thermodynamic muscle model and a chemical basis for A.V. Hill's muscle equation. *J Muscle Res Cell Motil.* **21**, 335–344 (2000).
5. J. E. Baker, I. Brust-Mascher, S. Ramachandran, L. E. LaConte, D. D. Thomas, A large and distinct rotation of the myosin light chain domain occurs upon muscle contraction. *Proc. Natl. Acad. Sci. U. S. A.* **95**, 2944–9 (1998).
6. J. E. Baker, C. Brosseau, P. B. Joel, D. M. Warshaw, The biochemical kinetics underlying actin movement generated by one and many skeletal muscle myosin molecules. *Biophys. J.* **82**, 2134–47 (2002).
7. J. T. Finer, R. M. Simmons, J. A. Spudich, Single myosin molecule mechanics: piconewton forces and nanometre steps. *Nature*. **368**, 113–119 (1994).
8. J. E. Baker, A chemical thermodynamic model of motor enzymes unifies chemical-Fx and powerstroke models. *Biophys. J.* **121**, 1184–1193 (2022).
9. D. M. Warshaw, W. H. Guilford, Y. Freyzon, E. Krementsova, K. a Palmiter, M. J. Tyska, J. E. Baker, K. M. Trybus, The light chain binding domain of expressed smooth muscle heavy meromyosin acts as a mechanical lever. *J. Biol. Chem.* **275**, 37167–72 (2000).
10. T. J. Stewart, V. Murthy, S. P. Dugan, J. E. Baker, Velocity of myosin-based actin sliding depends on attachment and detachment kinetics and reaches a maximum when myosin binding sites on actin saturate. *J. Biol. Chem.* **297**, 101178 (2021).
11. J. E. Baker, D. D. Thomas, Thermodynamics and kinetics of a molecular motor ensemble. *Biophys. J.* **79**, 1731–6 (2000).
12. A. F. Huxley, Muscle structure and theories of contraction. *Prog. Biophys. Biophys. Chem.* **7**, 255–318 (1957).
13. T. L. Hill, Theoretical formalism for the sliding filament model of contraction of striated muscle. Part I. *Prog. Biophys. Mol. Biol.* **28**, 267–340 (1974).

**Acknowledgments:** I thank JWG, LB, Julie, my students, colleagues, and mentors who have over many years inspired and guided this work. This was funded by a grant from the National Institutes of Health 1R01HL090938-01.

**Funding:** JEB was funded by a grant from the National Institutes of Health 1R01HL090938.

**Competing interests:** Author declares that they have no competing interests.

**Data and materials availability:** All data are available in the main text.

**Fig. 1. Entropy of mixing in a two-state chemical model.** (A) A chemical scheme shows a molecule that isomerizes with forward, $f_+$, and reverse, $f_-$, rates between two states that differ only in color. State B is blue, and state Y is yellow. (B) At t = 0, a closed system contains 10 such molecules all in state B (left panel). With a relaxation time constant, $\tau$, the entropic contribution to the free energy for the reaction in panel A irreversibly (single right arrow) drives the system to a state characterized by an equilibrium mixture of states B and Y (right panel). (C) In a bulk solution, the reaction in panel B appears as a solution that irreversibly changes color from blue to green.

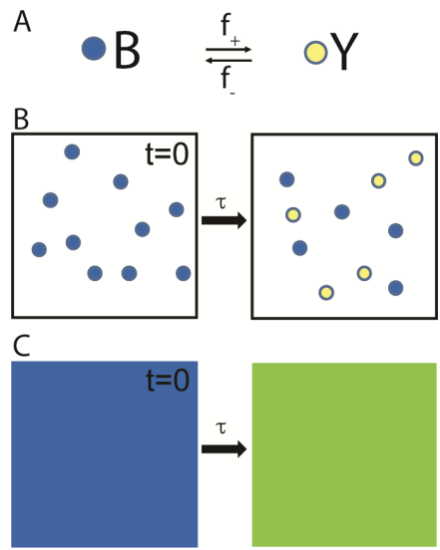

**Fig. 2. Entropy of mixing in a binary mechanical system.** (A) A chemical scheme shows a molecule that isomerizes with forward, $f_+$, and reverse, $f_-$, rates between two states that differ by a mechanical displacement, $d$. State B is a myosin detached from actin. State Y is a myosin bound to actin. The transition from state B to Y displaces actin relative to myosin. (B) At t = 0, a closed binary mechanical system contains 10 such molecules all in state B (left panel). With a relaxation time constant, $\tau$, the entropic contribution to the free energy for the reaction in panel A (single right arrow) drives the system to a state characterized by an equilibrium mixture of states B and Y (right panel). The net increase in the number of molecules in state Y results in a net displacement of the actin filament (attached to a freely movable wall) relative to myosin (attached to a fixed wall). (C) An equilibrium binary mechanical system in state [5,5] at $F = 0$ (panel B, right) is pulled in a direction that reverses the displacement of the actin filament in panel B, generating force $\Delta F_{ext} = -\Delta F_1$ in a system spring of stiffness $\kappa_{sys}$. The system responds with an average transition of one molecule from Y to B that reverses $\Delta F_{ext}$ resulting in $F = 0$. This is the case if no entropic force is generated with the same step.

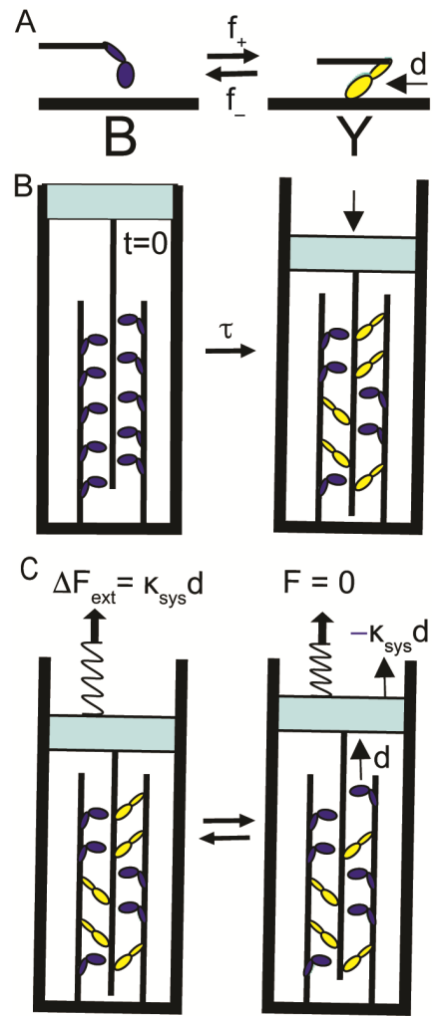

**Fig. 3. Forces required to unmix a binary mechanical system ($\kappa_{sys}$ = 0.125 pN/nm and $d$ = 4 nm).** (A) A binary system like that in Fig. 2C only with $N$ = 11 molecules is pulled to generate the force, $\Delta F_{ext}$, required to unmix the system from equilibrium state [8,3] to [9,2]. The system responds with a decrease in mechanical force, $\Delta F_1 = -\kappa_{sys}d$ (blue arrow) and an increase in entropic force, $\Delta F_S = T\Delta\Delta S/d$ (red arrow) associated with that step. The overall transition starts and ends along the isotherm (Eq. 4, red line) (B) A series of unmixing steps like that in panel A illustrates how mixing stalls (asterisk) when the finite molecular driving force $-\kappa_{sys}d$ (maroon bar) equals the entropic resistive force $T\Delta\Delta S/d$ (horizontal dashed lines).

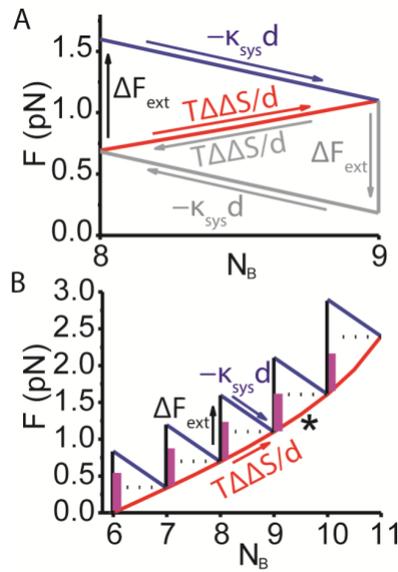